\documentclass[doublecol]{epl2}

\usepackage{graphicx}
\usepackage{float}
\usepackage{amsmath, amssymb}
\usepackage{lmodern}
\usepackage[T1]{fontenc}
\usepackage[utf8]{inputenc}
\usepackage[english]{babel}
\usepackage{color}
\usepackage{hyperref}

\title{Model for clustering of living species}
\shorttitle{Model for clustering of living species}

\author{D. Bazeia\inst{1} \and M.V. de Moraes\inst{2} \and B.F. de Oliveira\inst{2}}
\shortauthor{D. Bazeia \etal}

\institute{                    
  \inst{1} Departamento de F\'\i sica, Universidade Federal da Para\'\i ba, 58051-970 Jo\~ao Pessoa, PB, Brazil\\
  \inst{2} Departamento de F\'\i sica, Universidade Estadual de Maring\'a, 87020-900 Maring\'a, PR, Brazil
}


\abstract{
Clusters appear in nature in a diversity of contexts, involving
distances as long as the cosmological ones, and down to atoms and
molecules and the very small nuclear size. They also appear in several
other scenarios, in particular in biological systems as in ants, bees,
birds, fishes, gnus and rats, for instance. Here we describe a model
composed of a set of female and male individuals that obeys simple
rules that rapidly transform an uniform initial state into a single
cluster that evolves in time as a stable dynamical structure. We show
that the center of mass of the structure moves as a random walk, and
that the size of the cluster engenders a power law behavior in terms
of the number of individuals in the system. Moreover, we also examine
other possibilities, in particular the case of two distinct species
that can evolve to form one or two distinct clusters.}

\begin{document}

\maketitle

\section{Introduction}

Clusters appear in nature in several distinct situations. At very
large scales, the Universe, for instance, is known to be arranged in
the form of clusters of galaxies, and the galaxies themselves are
clusters of gas, dust, stars and their planetary systems; see, e.g.,
Ref. \cite{bookG}. At very small scales, at the nuclear level, for
instance, the nuclear matter aggregates to form the atomic nuclei,
which may also contain clusters of deuterons, tritons and alpha
particles; see, e.g., Ref. \cite{bookN}. As one knows, at the atomic
and molecular level, nuclei and electrons also aggregate to form atoms
and molecules. And more, clusters also appear in several other
scenarios: in economics, they may engender a strategy to improve
productivity \cite{eco}; in social networks, they may help control
specific features such as disease spreading \cite{social}; and they
may also play a role in data control and mining \cite{control,mining}
and in several other areas of research of current interest.

We may say that clusters are also of interest in living systems such
as ants, bees, birds and fishes, among others; see, e.g., Refs.
\cite{bookEco,book,Sci,book2} and references therein. In ants and
bees, the individuals tend to cluster in the presence of at least one
distinct individual, but this is not the case for birds and fishes,
for instance, since each species seems to have no particular
individual to lead or conduct the group. In this sense, the study of
clusters and the investigation of clustering mechanisms is of great
interest in science in general and, in particular, in the case of
living systems. As one knows, living in groups may engender the
advantage of a lower predation risk and better efficiency when seeking
for food, but may also enhance competition for food itself, and
increase the risk of illness due to the spread of diseases, among
other things; see, e.g., Refs. \cite{A,B,C,D,E,F} and references
therein for recent studies on the subject.

In this work we concentrate on the presence of clusters in a simple
system composed of a set of female and male individuals of a single
species that can move, and die or reproduce at a given rate. We
introduce no preference or distinction among the many individuals, so
the system is somehow closer to birds, cockroaches, fishes, gnus or
rats, for instance. We shall deal with a two-dimensional distribution
of individuals in a square box, so the model will be closer to
cockroaches, gnus or rats, among other possibilities. We also notice
the occurrence of sexual processes in bacteria such as the {\it
Escherichia Coli} \cite{natureB,conjug}, so the model may be of
interest at the bacterial level as well.

In order to examine issues related to the presence of clusters in
living systems, we start the work with a two-dimensional system of
individuals, searching for a mechanism that leads to the clustering of
the individuals into a state that is dynamically stable. We describe
the system in the Sec. Model and study its time evolution in Sec.
Results, where one shows the results for the cluster formation, and
some of its features. We examine other interesting possibilities in
Sec. Other Results and end the work with some conclusions and open
questions.

\section{Model}
\label{sec:model}

The model to be investigated is initially composed of a set of $N$
individuals, with $N/2$ being female and $N/2$ male. In this work we
only deal with species in which the individuals are either female or
male for their entire lifetime, with the initial state being prepared
with the total number of individuals distributed randomly in a square
box of linear size $L=1$. The spatial distribution of individuals is
described as an off-lattice model, in a way similar to the off-lattice
model considered before in \cite{PRL,2018-Avelino-EPL-121-48003}.
In other words, the space is continuum here, in distinction to the
discrete lattice model which will not be considered in this work. Due
to the random distribution, the initial state represents an uniform
distribution of individuals. The system obeys periodic boundary
conditions, and the time evolution starts with a time step in which
one randomly selects an individual, which is the active individual. It
is then moved with a fixed step defined by the distance $\ell=0.01$,
in a direction that is chosen randomly. After the motion, an action is
selected with the following probabilities: $p_r$ or $p_d$, for the
individual to reproduce or die, respectively. They are associated to
reproduction or extermination, and the real world is of course much
more complex than this, but here $p_d$ is used to describe predation 
on general grounds, and $p_r$ on the contrary is to keep the system
alive.

When $p_d$ is chosen, the active individual is removed from the
system. However, when $p_r$ is chosen, one verifies if there is an
individual of the opposite sex is the region inside the circle of
radius $\ell$ around the active individual; if there is no individual
we restart choosing another active individual, but if there are more
than one individual of the opposite sex in the region one chooses the
closest one. The next step is to check the total number of
individuals: if it is less than $N$, a new individual is born, which
is chosen to be female or male with equal probability. This individual
is put inside the box, at the distance $\ell$ of the female, in a
direction which is also chosen randomly. We have to choose $p_d<p_r$,
and here we take $p_d=0.3$ and $p_r=0.7$. This choice favours
reproduction, so to avoid an indefinite increasing in the number of
individuals, we add the constraint that the total number of
individuals should not overcome $N$.

We notice that the proximity parameter $\ell$ is introduced
to unveil the scale of reproduction. It is responsible for the
grouping of female and male into the cluster and describes a very
simple mathematical model, capable of conducting complex collective
phenomena based on very simple mathematical rules. Since no other
specific biological characteristics is attributed to the individuals
in the model, we cannot use it to describe any specific group of
living species. However, it will trigger a mechanism that leads to a
complex clustering phenomenon, which will be further explored in this
work. The motivation here is similar to the ones included in Refs.
\cite{B,C,AA, BB, CC, DD}, for instance, which also deal with simple
mathematical models that may be able to describe complex behaviors in
evolutionary game theory. We recall, in particular, the review on the
study of evolutionary dynamics of group interactions on structured
populations, complex networks and coevolutionary models \cite{B},
results showing that smart and tolerant species have more efficient
networks \cite{C}, the investigation where specific types of
reciprocity norms may lead individuals to split into groups in which
they are cooperative \cite{BB}, the study of topological frustration
on the evolutionary dynamics of the snowdrift game on a triangular
lattice \cite{CC}, and also the recent review on human cooperation,
with focus on spatial pattern formation, on the time evolution of
observed solutions, and on other behaviors that may either promote or
hinder socially favorable states \cite{DD}.

In this work, we run the numerical simulations using $t$ to provide the time
evolution in terms of generations, with one generation being the time
spent during the occurrence of $N$ time steps. We notice that the
square box of linear size $L$ describes a two-dimensional off-lattice
model, and we understand the distance $\ell$ as the parameter that
sets the scale of proximity between partners in the system. In
particular, we notice that $p_d$ occurs by chance, so it cannot be
seen as a behavioural characteristic of the individuals. However,
$p_r$ only occurs in the presence of the closest partner, inside the
circular region of radius $\ell$, if the total number of individuals
is below $N$, so the proximity parameter $\ell$ defines the
behavioural zone and acts to make $p_r$ a behavioural parameter. As we
shall show in Sec. Results, the above rules together favour
the extermination of isolated individuals and the grouping of partners
into a cluster that collects all the individuals of the system.

\section{Results}
\label{sec:results}

\begin{figure}[!htb]
\centering
\includegraphics[width=8.4cm]{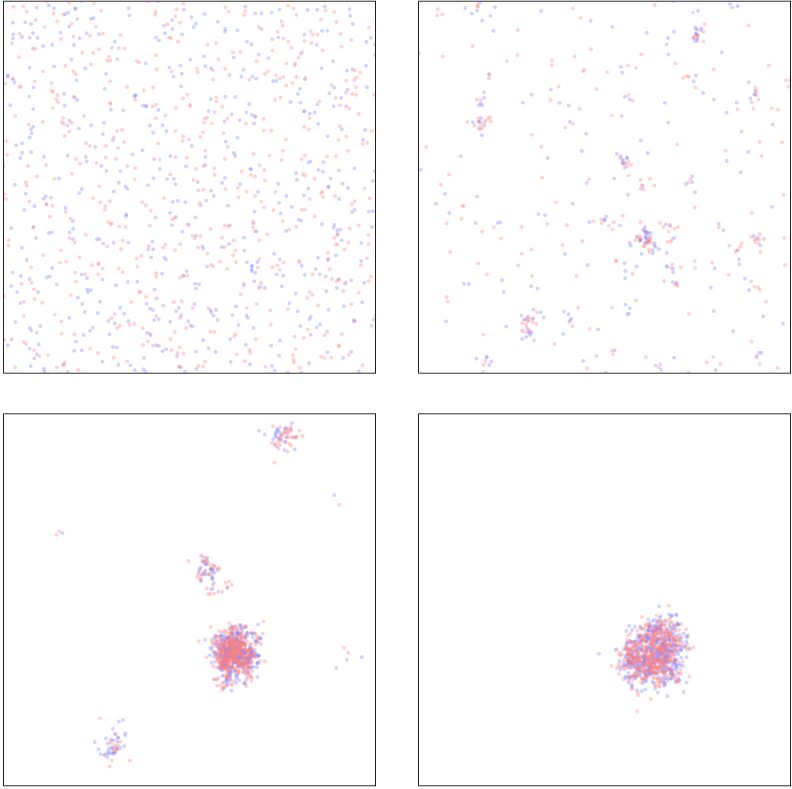}
\caption{Snapshots of the spatial distribution of females (red) and
males (blue) at the generation times $t=0$ (top left), $t=5$ (top
right), $t=25$ (bottom left), and $t=125$ (bottom right). One notices
the formation of clusters, with only one cluster surviving when time
goes beyond one hundred generations.}\label{fig1}
\end{figure}

We have implemented many numerical simulations, and in Fig. \ref{fig1}
four distinct snapshots are depicted to illustrate the time evolution
of the system with $N=1000$ individuals. The first snapshot at the top
left represents a typical initial state of the system. In this figure,
it is possible to verify that after some generations the system starts
to form small clusters, which evolve into larger ones, in an evolution
that ends up with a single larger cluster containing all the $N$
individuals present in the initial state. We can understand this with
the fact that in a larger cluster, the probability to implement the
rule reproduction increases due to the increasing of individuals in
the cluster, when compared to smaller clusters. The rules used to
describe the time evolution of the system seems to describe a very
efficient algorithm to cluster or aggregate the individuals into a
small region inside the box, so we move on to study the basic
properties of the system.

% figure %%%%%%%
% figure %%%%%%%

\begin{figure}[!htb]
	\centering
	\includegraphics[width=8.4cm]{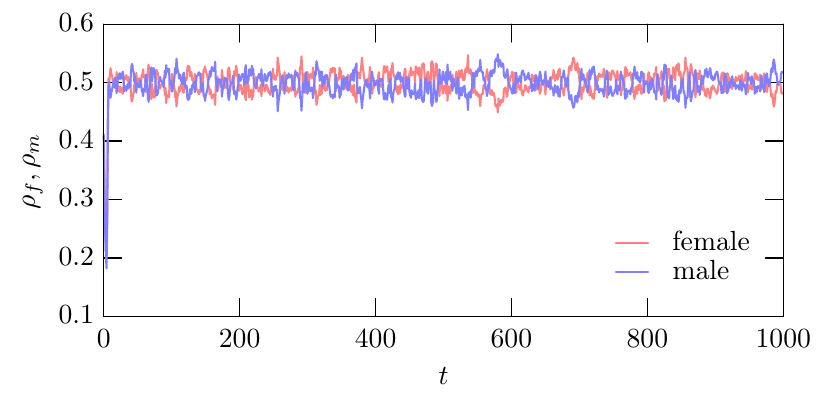}
	\caption{The female (red) and male (blue) abundances are displayed
as a function of time for a long time.}
	\label{fig2}
\end{figure}

In Fig. \ref{fig2} we examine the time evolution of the abundance or
density of female ($\rho_f(t)$) and male $(\rho_m(t))$ individuals. We
run a simulation using $N=1000$ up to $t=1000$ generations and depict
the results with the red and blue colors that represent the female and
male, respectively. The results show that they evolve fluctuating
around their average values
$\langle{\rho}_f\rangle=\langle{\rho}_m\rangle=0.5$ for very long
times, indicating the dynamical robustness of the time evolution of
the system. In fact, we tested the abundance for much longer times,
for $t=10000$, $20000$, $30000$, $40000$ and $50000$ generations, and
we found no important deviation from the behavior displayed in Fig.
\ref{fig2}. This suggests that the system rapidly evolves to reach
dynamical stability. However, we noticed that the simulation started
with a very large and abrupt variation of abundance, but we can
understand this as follows: initially, the individuals are on average
well separated from each other, favouring that the ratio of dead
overcomes reproduction, but this is soon modified due to the
clustering mechanism and the system rapidly returns to its dynamical
stability, with the abundances oscillating around their average with
small fluctuations. Female and male evolve on equal footing, so they
have similar behavior and the very same average abundance.

\begin{figure}[!htb]
	\centering
	\includegraphics[width=8.4cm]{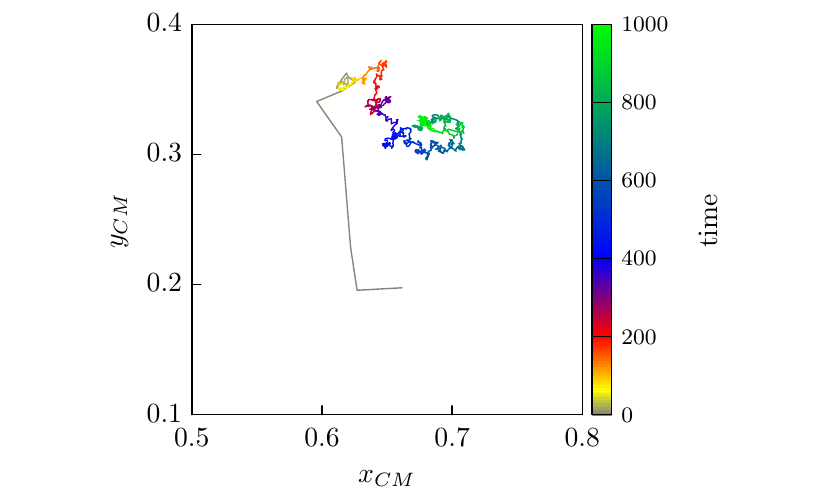}
	\caption{The position of the center of mass is shown as a functions
of time.}
	\label{fig3}
\end{figure}

The results displayed in Figs. \ref{fig1} and \ref{fig2} show that the
system rapidly evolves grouping all the individuals into a single
cluster, which seems to evolve robustly for very long times. For this
reason, let us now concentrate on the behavior of the cluster that is
formed as the final state of the system. The first step concerned the
calculation of the {\it center of mass} of the system, supposing that
all the individuals carry the same mass. The calculation follows the
method suggested in \cite{2008-Linge-JGT-13-53}, which takes into
account the periodic boundary conditions that we are using in this
work. Since the system evolves in time, the center of mass coordinates
are $(x_{CM}(t), y_{CM}(t))$, and follows the trajectory displayed in
Fig. \ref{fig3}, colored in accordance with the time evolution that
appears at the right of the figure. After a hundred generations the
cluster is generated, and then its center of mass position evolves
behaving like a random walk, illustrated by the colors yellow, red,
blue and green in the figure. We notice, however, that the center of
mass moves rapidly at the beginning of the simulation, before the
formation of the cluster.

% figure %%%%%%%
% figure %%%%%%%

% figure %%%%%%%
% figure %%%%%%%

\begin{figure}[!htb]
	\centering
	\includegraphics[width=8.4cm]{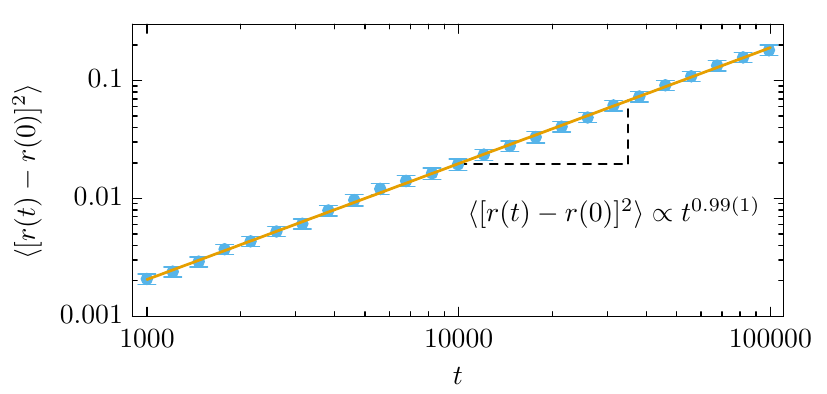}
	\caption{The center of mass mean square displacement of the cluster
as a function of time for a very long time.}
	\label{fig4}
\end{figure}

We remark that the two Figs. \ref{fig2} and \ref{fig3} were
constructed using results of a single realization, in fact the very
same realization used to depict Fig. \ref{fig1}. Moreover, to certify
that the center of mass position shown in Fig. \ref{fig3} moves like a
random walk for larger values of the time evolution, we have depicted
in Fig. \ref{fig4} the center of mass mean squared displacement (MSD),
which was calculated as follows: we first discard the $1000$ first
positions of the center of mass, to ensure that the cluster is already
formed, then we start counting the time and at $t$ we calculate the
MSD from the $1000$ center of mass positions, since we are simulating
the system $1000$ times. We do this for several values of $t$ in order
to calculate the mean square displacement. We call it
$\langle[r(t)-r(0)]^2 \rangle $ and depict the results with the light
blue dots in the figure. The results shows that the MSD varies
linearly on time which is characteristic of Brownian motion \cite{PR}.
The error is due to the fitting procedure. The bins in Fig. \ref{fig4}
represent the error bars that account for the procedure to calculate
the numerical values. We notice that the linear behavior displayed in
Fig. \ref{fig4} persists for a very long time, up to $100000$
generations, once again indicating the dynamical stability of the
cluster.

% figure %%%%%%%
% figure %%%%%%%

% figure %%%%%%%
\begin{figure}[!htb]
	\centering
	\includegraphics[width=8.4cm]{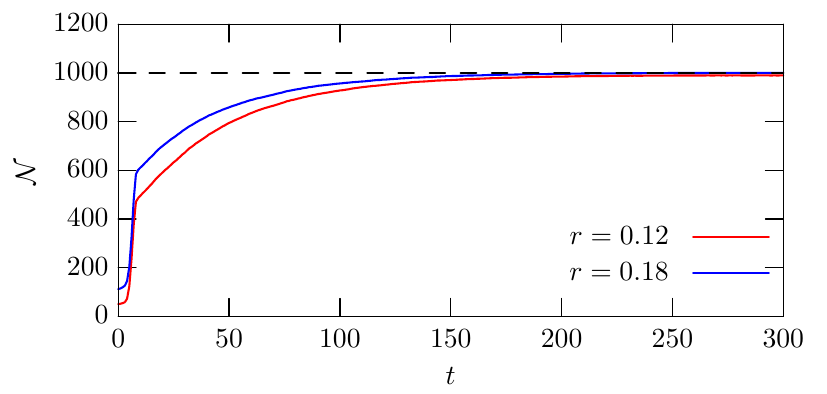}
	\caption{Number of individuals inside a circle of radius $r$
	and center at the center of mass position of the system, depicted for $r=0.12$ and $0.18$.}
	\label{fig5}
\end{figure}
% figure %%%%%%%

The appearance of very large fluctuations at early times, as shown in
Fig. \ref{fig2}, and the identification in Fig. \ref{fig3} that the
center of mass moves very rapidly at the beginning of the simulations
has driven our attention to double check the numerical simulations. In
particular, we display in Fig. \ref{fig5} the number ${\mathcal N}$ of
individuals inside a circle of radius $r$ and center at the center of
mass position of the system as a function of time. The figure is
depicted for $N=1000$ individuals, with $r=0.12$ (red) and $0.18$
(blue), with the data coming from an average over $1000$ simulations.
The values of $r$ are chosen to lead to circles that fill $5$ and $10$
percent of the total area of the square box of linear size $L=1$,
respectively. As expected, the initial evolution is abrupt, but the
system rapidly evolves to approach a smooth time evolution which
ultimately leads to the cluster formation. We also notice from Fig.
\ref{fig5} that the red and blue curves behave very similarly, and
that almost all of the individuals are inside the smaller circle when
$t$ approaches 200 generations.

\begin{figure}[!htb]
	\centering
	\includegraphics[width=8.4cm]{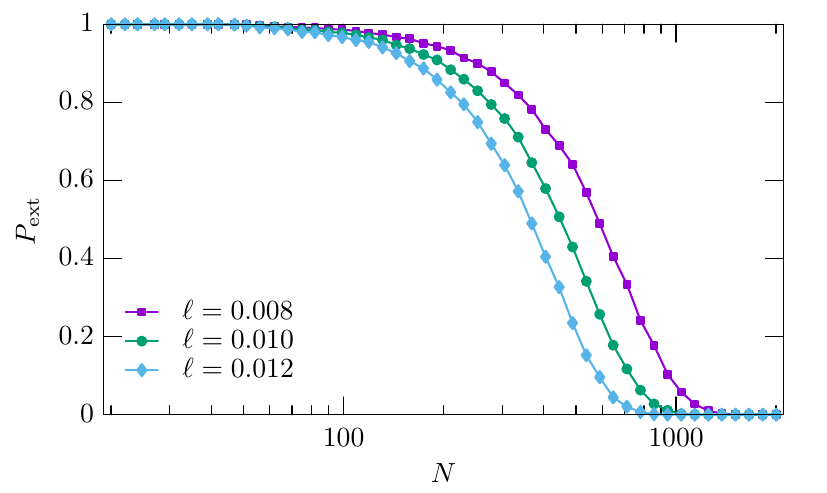}
	\caption{Extinction probability as a function of the number of
individuals, depicted for $\ell=0.008$, $0.01$ and $0.012$.}
	\label{fig6}
\end{figure}

We further noticed that if the density of individuals at the
initial state is sufficiently small, the average distance among
individuals may be sufficiently large to make the ratio of dead
overcome reproduction with no return, that is, conducting the system
to extinction. To examine this issue appropriately, we have studied
the extinction probability $P_{\textrm{ext}}$ as a function of the
number of individuals $N$ for many distinct values of $N$, always
starting with an initial state which is constructed randomly, as
described in Sec. Model. The results are depicted in Fig. \ref{fig6}
for three distinct values of $\ell$, to show how the proximity
parameter changes the behavior of the system. We notice that the
extinction probability vanishes and the system evolves in time keeping
coexistence between female and male for appropriate values of $\ell$
and $N$. However, as we suspected, the probability of extinction
increases as one decreases the number of individuals. Each dot
displayed in Fig. \ref{fig6} is calculated as the average in a set of
$10000$ simulations, with all the simulations evolved for $1000$
generations. At the end of each simulation, we verified for the
vanishing of individuals, or the presence of individuals of a single
sex since this also implies extinction. The results displayed in Fig.
\ref{fig6} show that for the box with unity linear size $L=1$, the use
of the distance $\ell=0.01$ is of interest if one takes $N=1000$ or
other higher values. This will be used below to describe other
features of the system.

% figure %%%%%%%
% figure %%%%%%%

\begin{figure}[!htb]
	\centering
	\includegraphics[width=8.4cm]{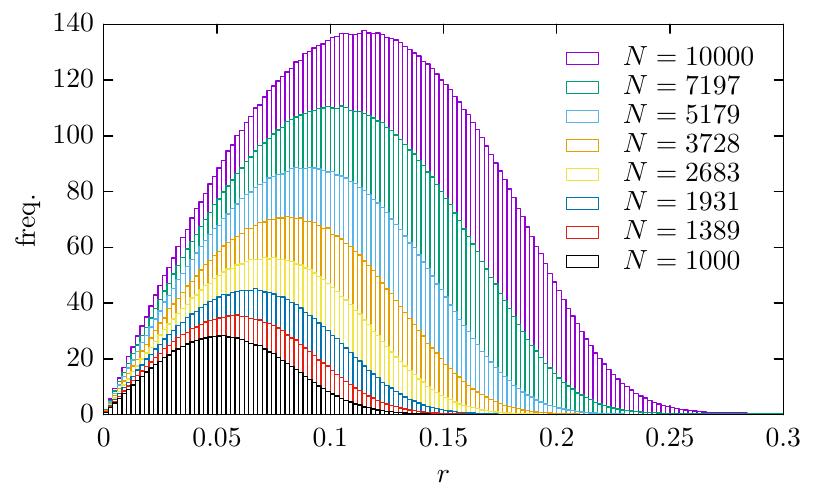}
	\caption{Histogram of the distribution of distance of the individuals to the center of mass of the cluster for several values of $N$.}
	\label{fig7}
\end{figure}

In order to further examine the cluster, we have studied the
distribution of distance of the individuals to the center of mass. The
Fig. \ref{fig7} shows this distribution for all the individuals. These
results are obtained for several values of $N$, and they are
calculated as an average of $1000$ distinct simulations for each value
of $N$, for $1000$ generations. 

\begin{figure}[!htb]
	\centering
	\includegraphics[width=8.4cm]{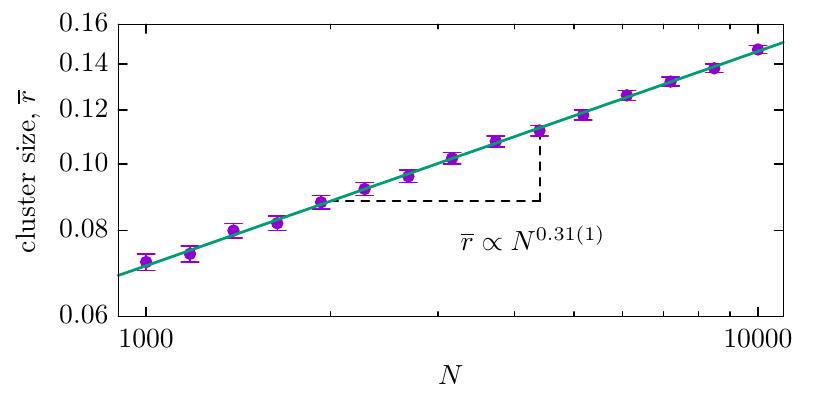}
	\caption{Cluster size as a functions of the number of individuals.}
	\label{fig8}
\end{figure}

With the results shown in Fig. \ref{fig7}, we could quantify the size
of the cluster in terms of the number of individuals, which is
depicted in Fig. \ref{fig8}. It is calculated as the width at half of
its maximum height, and the results nicely show a power law behavior,
that is, the size of the cluster is proportional to the number of
individuals to the power $\alpha$, in the form $\bar{r}\propto
N^{\alpha}$, with $\alpha=0.31(1)$, with the error in $\alpha$ being
dictated by the fit. The error bars in Fig. \ref{fig8} are due to the
numerical simulations and are represented by the size of the bins in
the figure.

% figure %%%%%%%
% figure %%%%%%%
\section{Other results}
\label{sec:other}

\begin{figure}[!htb]
	\centering
	\includegraphics[width=8.4cm]{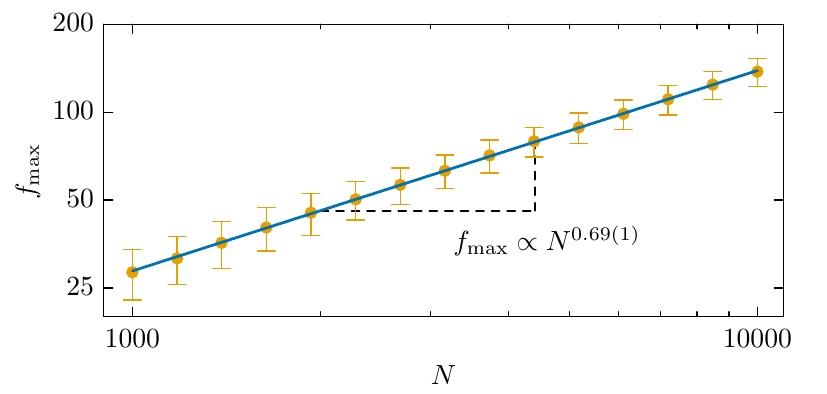}
	\caption{The highest frequency as a functions of the number of individuals.}
	\label{fig9}
\end{figure}

Let us now examine the above model under other conditions. We first
considered other values of $p_r $ and $p_d $; in particular, we used
$p_r=0.8 $ and $p_d=0.2 $ and $ p_r=0.9$ and $p_d=0.1 $ and noticed no
important qualitative difference from the results described above. We
also observed that in the histograms displayed in Fig. \ref{fig7}, the
distribution of frequency follows an interesting pattern, and this
instigated us to study the issue more accurately. We first noticed
that the highest frequency depends on the number of individuals, so we
studied this to depict in Fig. \ref{fig9} the function $f_{\rm
max}(N)$, which represents the highest frequency in terms of the
number of individuals. Interestingly, the results show that it depends
on $N$ in the form of a power law, with power $0.69(1)$.

The above model considers female and male on an equal footing.
However, since sex ratio varies widely in nature we can choose other
possibilities; see, e.g., Ref.~\cite{nature} and references therein.
We first considered the same rules, but started with the initial state
with female and male unevenly distributed. We used several
possibilities, such as $0.6$ and $0.4$, $0.7$ and $0.3$, and $0.8$ and
$0.2$, for female and male, for instance. In all cases the system
always relax to a single cluster with equal distribution of female and
male individuals. We also investigated the case where reproduction
evolves under the male-biased rate of $0.55$; that is, when an
individual is selected to be born, there is higher $55\%$ chance that
a male will be born. We examined the numerical simulation in this
case, and found that the system also develop a single cluster, but now
keeping the very same bias: the abundance fluctuates around $55\%$
male and $45\%$ female. Alternatively, we kept the same rule of the
previous Sec. Model for reproduction, but we changed $p_d$ as follows:
when $p_d$ is selected, it is effectively implemented with the rate $
50\%$ if the individual is a male, and with the full rate $100\% $ in
the case of a female. The results showed that the system evolves to
form a cluster, but now with the abundance of female and male
fluctuating around $1/3$ and $2/3$, respectively. These results show
that if there is no sex-biased rule, the system evolves to form a
cluster with female and male evenly distributed; however, when a
sex-biased rule is present, the system also form a cluster, but now
with female and male distributed under the same bias.

\begin{figure}[!htb]
	\centering
	\includegraphics[width=8.4cm]{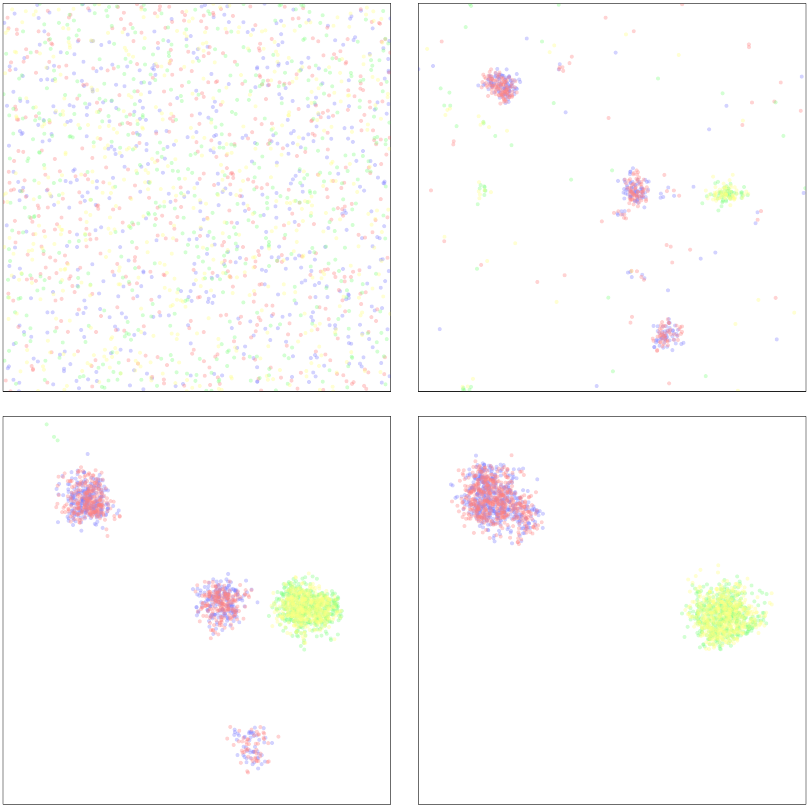}
	\caption{Snapshots of the spatial distribution of females
(red/yellow) and males (blue/green) at the generation times $t=0$ (top
left), $t=5$ (top right), $t=25$ (bottom left), and $t=125$ (bottom
right). One notices the formation of clusters, with only two clusters
surviving when time goes beyond one hundred generations.}
	\label{fig10}
\end{figure}

We have also enlarged the system to consider two distinct species, one
of them being the $1000$ red and blue individuals, and the other
$1000$ yellow and green individuals. We suppose that they live
together in a square box of unity linear size, and obey the very same
rules described in Sec. Model, but now we add another constraint in
the reproduction: any individual can only reproduce if inside the area
with $\ell=0.01$, there is no individual of the other species. This
adds a repulsion between the two species, which contributes to the
formation of two distinct and spatially separated clusters, that
evolve independently, with the very same characteristics that we have
already identified in Sec. Results. Some results are displayed in Fig.
\ref{fig10}, where we show the initial state at $t=0$, and its
evolution at $t=5$, $25$, and $125$. We compare this with Fig.
\ref{fig1} to see that the system now relaxes to two distinct spatial
clusters that evolve as two stable structures, independent from one
another. The two distinct clusters are formed when we keep the initial
number of individuals in each species as its upper bound for
reproduction; however, if we choose the initial total number of
individuals in the two species as the upper bound for reproduction,
the system always relax to a single cluster with all the individuals
of the same species, exterminating the other species. The two
situations are quite distinct: the first case may be more appropriate
to describe two distinct species that have independent constraints for
reproduction; the second case is different, and is more appropriate to
describe two distinct species that have the same constraint for
reproduction.

% figure %%%%%%%
% figure %%%%%%%

% figure %%%%%%%
% figure %%%%%%%

% figure %%%%%%%
% figure %%%%%%%

\section{Conclusion}
\label{sec:end}

In this work we investigated a simple model that describes a set of
$N$ female and male individuals that are arranged in a off-lattice
square box of linear size $L$. The individuals may die or reproduce,
with the reproduction only occurring if the partner individual is very
close to the active individual. We have used appropriate values to
describe the number of individuals, the size of the box and the
partner proximity, and to quantify the probabilities to die or
reproduce. We run the numerical simulations for very long times, and
the results suggested that the system rapidly evolves into a cluster
that is dynamically stable. In particular, we calculated the
abundances of female and male, and found that they evolve similarly,
fluctuating around the same average as time goes by.

We also calculated the center of mass position of the cluster, which
showed a behavior that approaches a random walk motion when one runs
the simulations for a long time. Since both the abundance and the
center of mass motion have peculiar behavior in the beginning of the
simulations, we also investigated the extinction probability of the
individuals as a function of the number of individuals, keeping the
box size $L$ fixed and using three distinct values of the proximity
parameter $\ell$. The general result here is that for a given $\ell$,
small values of $N$ may drive the system to extinction, but we have a
lot of room to choose $N$ to keep the system evolving from the uniform
initial state to a cluster which is dynamically stable. The size of
the cluster was also investigated for $L$ and $\ell$ fixed. We
calculated the distribution of distance of the individuals to the
center of mass, from which we obtained the mean square displacement of
the cluster. The results unveiled a dependence on $N$ that follows a
power law behavior.

We have also considered some modifications in the initial state and on
the rules. In particular, if we keep the same rules and consider an
initial state with female and male unevenly distributed, we end up
with a stable cluster composed of female and male evenly distributed
inside the structure. Also, if we consider sex-biased rules, the
cluster is also formed and is also dynamically stable, but it now
keeps female and male with the same bias provided by the sex-biased
modification. We have also considered two distinct species, adding a
very simple modification in the rule of reproduction. The modification
may make the system to evolve with the formation of two distinct
clusters, one for each species, or, else, a single cluster composed of
only one of the two species. The two cases are of current interest,
and the case with the formation of two independent clusters allows
that we examine coexistence of two distinct species, which can also be
extended to several species.

Since the clustering mechanism used in this work is simple, we may add
other more sophisticated rules to help us understand specific features
of clusters of living systems. We can, in particular, consider the
case of groups of simple organisms that reproduces using the doubling
mechanism, and also of groups of several distinct individuals that
interact with one another, among other possibilities. We can also
change the behavioural rule for reproduction, modifying the way it
acts in the area defined by $\ell$, to get to other dynamically stable
grouping states. Another issue of current interest concerns the use of
the off-lattice model in a cube, instead of the square box considered
in this work, to contribute to describe the behavior of clusters in
space. These and other related issues are now under consideration and
we hope to report of them in the near future.

\acknowledgments
We would like to thank I.M. Medri for discussions. The study was
financed in part by Conselho Nacional de Desenvolvimento Cient\'\i
fico e Tecn\'ologico (CNPq) and Coordenação de Aperfeiçoamento de
Pessoal de Nível Superior (CAPES, Finance Code 001). B.F.O. thanks
Funda\c c\~ao Arauc\'aria and INCT-FCx (CNPq/FAPESP) for financial and
computational support. D.B. thanks CNPq (Grant No. 06614/2014-6) and
Para\'\i ba State Research Foundation (FAPESQ-PB, Grant No.
0015/2019).

\end{document}